\documentclass[%
reprint,
superscriptaddress,
showpacs,preprintnumbers,
amsmath,amssymb,
aps,
]{revtex4-1}

\usepackage{color}
\usepackage{graphicx}% Include figure files
\usepackage{dcolumn}% Align table columns on decimal point
\usepackage{bm}% bold math
\begin {document}
%section {title}
%\preprint{APS/123-QED}

\title{Aging generates regular motions in weakly chaotic systems}% Force line breaks with \\
%\thanks{A footnote to the article title}%

\author{Takuma Akimoto}
\email{akimoto@z8.keio.jp}
\affiliation{%
  Department of Mechanical Engineering, Keio University, Yokohama, 223-8522, Japan
}%

\author{Eli Barkai}
%\email{akimoto@z8.keio.jp}
\affiliation{Physics Department, Bar-Ilan University, Ramat-Gan 52900, Israel
}%

%\collaboration{MUSO Collaboration}%\noaffiliation

\date{\today}% It is always \today, today,
%  but any date may be explicitly specified

\begin{abstract}
Using intermittent maps with infinite invariant measures, 
we investigate the universality of time-averaged observables under
 aging conditions. 
According to Aaronson-Darling-Kac theorem, 
in non-aged dynamical systems
 with infinite invariant measures, 
the distribution of the normalized time averages of integrable functions 
converge to the Mittag-Leffler distribution. 
This well known theorem holds when the start of observations
coincides with the start of the dynamical processes. 
Introducing a concept of the aging limit where the aging time $t_a$ and the total measurement 
time $t$ goes to infinity while the aging ratio $t_a/t$ is a constant, 
we obtain a novel distributional limit theorem of  time-averaged observables integrable with respect
to the infinite invariant density.   
Applying the theorem to the Lyapunov exponent in 
intermittent maps, we find that regular motions
 and a weakly chaotic behavior coexist in the aging limit. 
This mixed type of dynamics is controlled by the aging ratio and hence
is  very
different from the usual scenario of regular and chaotic motions in Hamiltonian systems.
The probability of finding regular motions in non-aged processes is
zero, while in the aging regime it is finite and it increases when
system ages. 
\end{abstract}

% PACS, the Physics and Astronomy
% Classification Scheme.
%\keywords{Suggested keywords}%Use showkeys class option if keyword
%display desired
\maketitle

%\tableofcontents

\section{Introduction}
Aging is a concept describing  slow relaxation phenomena in spin glasses \cite{Bouchaud1992}, %glasses, 
 interface fluctuations in liquid-crystal turbulence \cite{Takeuchi2012}, blinking quantum dots \cite{Brok2003, Margolin2004}
and transports in cells \cite{Weigel2011}. 
In nature dynamical processes may start at time $-t_a$ long before
the actual measurement of the process begins at $t=0$. 
In aging systems, statistical quantities measured in the time interval $[0,t]$ 
are crucially affected by the 
aging time $t_a$ \cite{Burov2010, Schulz2012}.  
For example the  distribution of time
averaged mean square displacement was considered in \cite{Schulz2012} using stochastic
tools relevant to diffusion of bio-molecules in the cell.  
In contrast, for stationary and ergodic processes, 
statistical properties of time-averaged observables do not depend on 
the aging time $t_a$ if 
the measurement time is large enough ($t\rightarrow\infty$). 
Here, we introduce the aging limit where the aging time $t_a$ and the  total measurement time $t$ goes to infinity 
and the ratio $T_a\equiv t_a/t$ (the aging ratio) is fixed as a constant. 
We proceed to show universal statistical properties 
of time-averaged observables in  a class of dynamical systems, extending infinite ergodic
theory \cite{Aaronson1997} to the aging regime. 

 Exponential separation of nearby trajectories, i.e., chaos is a feature
found in many dynamical systems. 
In many generic Hamiltonian systems, the phase space is either chaotic or 
 regular. However, closer look at dynamics many times
reveals a mixed phase space. This means that parts of the trajectories
are rather regular, for example Kolmogorov-Arnold-Moser tori
in phase space, while others are chaotic \cite{Aizawa1989a}.
 Determining which type of motion depends ultimately on the choice of initial conditions.
Here we investigate a completely new mechanism for dual structure
of dynamics. We show how, for a class of dynamical systems possessing
an infinite invariant measure, dynamics is generically split into regular
and weakly chaotic. This splitting is controlled by the age of the process,
and hence is completely different from usual scenario.

 In weakly chaotic systems the separation of nearby trajectories 
is sub-exponential \cite{Gaspard1988}. In many cases such systems have an infinite 
invariant density \cite{Akimoto2010a}, i.e., a non-normalizable density (see details below).
 Since chaos is a pre-condition of ordinary statistical mechanics, it is not
surprising that usual statistical concepts like ergodicity and stationarity
break down when treating weakly chaotic
systems. 
It is known that some dynamical systems with infinite invariant measures show an aging behavior \cite{Barkai2003}. 
Here we consider a dynamical system whose evolution
started at time $-t_a$ with $t_a>0$, while an observation of the
dynamical process starts at time $t=0$. Within the observation
time $(0,t)$ an observer evaluates time averages of observation functions in dynamical systems,
and we are interested in the ergodic properties of these time averages.
From the viewpoint of dynamical systems, aging means that a density at $t=0$ strongly depends on the aging time  $t_a$. 
If a dynamical system has an invariant {\it probability} 
measure, an initial density denoted by $\rho(x,t_a)$
  converges to an invariant density as $t_a\rightarrow \infty$, indicating that the aging ratio 
does not affect statistical properties of time-average observables. However, in an infinite measure system, 
$\rho(x,t_a)$ does not converge to an invariant measure as $t_a\rightarrow\infty$.

Ergodic theory states that the time averages of integrable functions converge to constant values (ensemble averages)
for almost all initial conditions if the dynamical system has an absolutely continuous invariant probability measure \cite{Birkhoff1931}. 
If an invariant measure cannot be normalized (infinite invariant measure), there does not exist an exponent $\alpha$ ($0<\alpha<1$) such that 
$\sum_{k=0}^{n-1}f (x_k)/n^{\alpha}$ converges to a non-trivial constant ($\ne 0$ and $\pm \infty$)
 for an integrable function $f(x)$ with respect to an invariant 
measure (see details below), where $\{ x_k\}_{k=0, \ldots }$ is a trajectory \cite{Aaronson1997}. 
That is, time averages of integrable functions do not converge to a constant. 
However, Aaronson's Darling-Kac (ADK) theorem \cite{Aaronson1981, Darling1957} states that time averages of integrable functions 
{\it converge in distribution}: there exists a sequence $|a_n| \propto n^{\alpha}$ such that 
\begin{equation}
\Pr \left( \frac{1}{a_n} \sum_{k=0}^{n-1} f(x_k) < x \right) 
\rightarrow \int_0^x d_{\alpha}(\xi)d\xi \quad {\rm as} \quad n\rightarrow \infty,
\label{ADK}
\end{equation}
where $d_{\alpha}(x)$ is the Mittag-Leffler density of order $\alpha$ \cite{Aaronson1997}.
In other words, the time average
depends strongly on an initial position but the distribution  of the time average converges to a universal distribution 
(the Mittag-Leffler distribution) for almost all smooth initial densities.  Here we investigate ergodic theory under the conditions of aging.
We show large differences from the ADK theorem in the aging regime
where both $t_a$ and $t$ are large, and  propose a limit
theorem describing distributions of  properly scaled sums $\sum_{k=0}^{\infty} f(x_k)$.

Recently, it was shown that infinite ergodic theory plays an important role in elucidating an intrinsic randomness 
of time-averaged observables in dichotomous processes modeling blinking quantum dots \cite{Akimoto2008} 
and anomalous diffusion \cite{Akimoto2010, Akimoto2012}. 
Since aging appears in infinite measure dynamical systems, it is an interesting and important problem to clarify 
whether distributional behaviors of time-averaged observables  
are affected by the aging ratio. Here, we provide  evidence that the 
aging limit where $t\to \infty$ and $t_a \to \infty$ but their ratio
$T_a\equiv t_a/t$ remaining finite,  plays a crucial role in characterizing 
behaviors of time-averaged observables. In particular,
we show that the distribution of time averages of integrable functions is determined by the aging ratio $T_a$
 using weakly chaotic systems.

\section{Aging dynamical systems}
We consider maps $T: [0,1]\rightarrow [0,1]$ which satisfy the following conditions for some 
$\gamma_1 \in (0,1)$: 
(i) the restrictions $T: (0,\gamma_1) \rightarrow (0,1)$ and $T: (\gamma_1,1) \rightarrow (0,1)$ are increasing, 
onto and $C^2$-extensions to the respective closed intervals; (ii) $T'(x)>1$ on $(0, \gamma_1] \cup [\gamma_1, 1]$; 
$T'(0)=1$; (iii) $T(x)-x$ is regularly varying at zero with index $1+1/\alpha$, $T(x)-x \sim a_0x^{1+1/\alpha}$ ($\alpha>0$). 
These maps are related to number theory \cite{Thaler1980}, intermittency \cite{Pomeau1980, Manneville1980, Aizawa1984}, and anomalous diffusions \cite{Geisel1984, Geisel1985, Zumofen1993, Artuso2003}. 
One of the best known examples is the Pomeau-Manneville map \cite{Pomeau1980, Manneville1980}: 
\begin{equation}
x_{t+1}=T(x_t)=x_t +  x_t^{1+1/\alpha}\quad {\rm mod} \hspace{.2cm} 1.
\label{map}
\end{equation}
In what follows, we use the map (\ref{map}) for numerical simulations. 
This famous map has a  marginally unstable
fixed point at
$x=0$ and hence trajectory is trapped in its vicinity
escaping slowly and then is reinjected back.
The map and its extensions have attracted wide interest since it exhibits
intermittency, power law distributed waiting times, weak chaos to name
a few novel effects.
According to Thaler's estimation \cite{Thaler1983},
an invariant density $\tilde{\rho}(x)$ is given by $\tilde{\rho}(x) \sim \tilde{h}(x) x^{-1/\alpha}$ for $x\in (0,1]$, 
where $\tilde{h}(x)$ is a positive  bounded continuous function on $[0,1]$.
 Notice that when $\alpha<1$ this invariant density cannot be normalized,
due to its divergence close to $x=0$. 
Here $\tilde{\rho}(x)$ is defined up to a multiplicative constant,
which we later specify. 
In what follows, we consider an infinite measure system ($0<\alpha<1$). In aging systems, the systems start  
at time $-t_a$ before the measurement is started at $t=0$, where $t_a$ is the aging time (see Fig. 1).
We assume that initial points $x_{-t_a}$ are uniformly distributed on $[0,1]$. 
The density $\rho_0(x)$ at $t=0$, when the measurement is started, is given by the density of $x_{t_a}$, i.e., 
$\rho(x,t_a)$.

\begin{figure}
\includegraphics[height=.65\linewidth, angle=0]{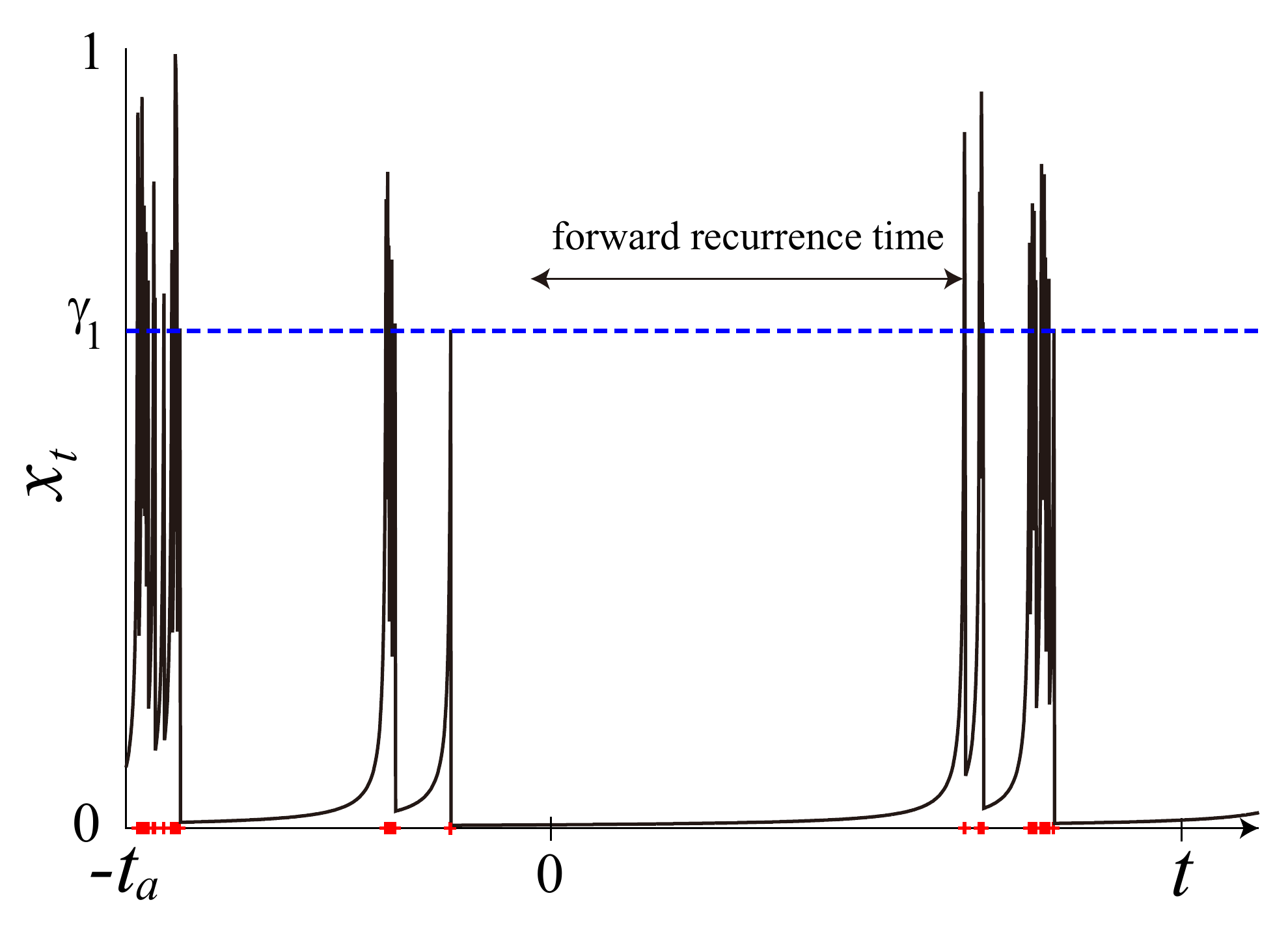}
\caption{Schematic view of an aging process in the map (\ref{map}) ($\alpha=0.8$). A trajectory is given by a solid line while 
renewals, $x_t \geq \gamma_1$, i.e., $\sigma(x_t)=1$, are depicted as red crosses on time axis.
$t=0$ is the time at which a measurement starts.}
\label{renewal_fig}
\end{figure}

\section{Renewal processes}
Renewal processes are point processes where interevent times of points are 
independent and identically distributed (IID) random variables \cite{Cox}. Therefore, renewal processes are 
characterized by the distribution of interevent times of renewals. 
Let us consider the following observation function 
\begin{equation}
\sigma(x) = \left\{
\begin{array}{ll}
0\quad & (x<\gamma_1)\\
\\
1 & (x\geq \gamma_1),
\end{array}
\right.
\end{equation}
where $\gamma_1<1$  attains $T(\gamma_1)=1$. We call the state $\sigma = 0$ the laminar phase while $\sigma =1 $ 
the chaotic phase. In the chaotic phase, trajectories $x_t$ show usual chaotic behavior  because of the condition (ii). Moreover, 
the jump transformation $T^{n(x)}$ on $x \in [\gamma_1,1]$, 
which is a transformation restricted to $[\gamma_1,1]$, has an absolutely continuous invariant probability 
measure whereas the original map $T$ does not, where $n(x) \equiv \min\{ k \geq 1: T^k(x) \in [\gamma_1,1]\}$ \cite{Thaler1983}. 
With the aid of these chaotic properties in the jump transformation,
trajectories $\sigma_t=\sigma(x_t)$ can be regarded as a renewal process because the interevent times between the events
$\sigma(x_t)=1$ are considered to be IID random variables. 
The probability density function (PDF) of the interevent times (or residence times of laminar phase) 
is given by \cite{Geisel1984}
\begin{equation}
\psi(\tau) \sim \alpha A^{\alpha} \tau^{-1-\alpha}, \quad \tau \rightarrow \infty,
\label{pdf_interevent_time}
\end{equation}
where the exponent of the PDF is controlled by the nonlinearity of the map in vicinity
of the indifferent fixed point through the maps parameter $\alpha$ and $A$ is a constant, which depends on the map. 
%We note that chaotic phases, $\sigma=1$, do not affect our main result (Theorem 1) at all. In other words, the map 
%restricted to $[\gamma_1,1]$ affect the PDF for small interevent times and the parameter $A$ but not affect the exponent 
%$\alpha$.

 Let $N_t$ be the number of renewals in the time interval $(0,t)$ for
a process which started on $t_a$, and with IID interevent times according to  Eq.~(\ref{pdf_interevent_time}). 
In  non-aged  renewal theory $t_a=0$, 
the distribution of the number of jumps $N_t/t^{\alpha}$ obeys the Mittag-Leffler distribution of order $\alpha$ ($<1$) 
\cite{Feller1971}. 
Here, we review a derivation of the distribution and extend it to the
aging regime \cite{Barkai2003b}. Let $S_n$ be the sum of the interevent times ($S_n=\tau_1 + \ldots + \tau_n$), 
then we have the following relation
\begin{equation}
\Pr  (N_t <n) = \Pr (S_n > t).
\end{equation}
We notice that
the interevent time PDF (\ref{pdf_interevent_time}) belongs to the domain of attraction of stable laws \cite{Feller1971}.
In what follows, we use the notation $\Pr_{t_a=0}(\cdot)$ for non-aged processes, and $\Pr(\cdot ; t_a)$ for aged processes. 
By the generalized central limit theorem and setting $n=t^{\alpha}x$, we have 
\begin{eqnarray}
\Pr_{t_a=0} (N_t/t^{\alpha}<x) &=& \Pr_{t_a=0} (S_n/n^{1/\alpha} > x^{-1/\alpha})\\
&\rightarrow& \int_{x^{-1/\alpha}}^{\infty} l_{\alpha}(y)dy \quad {\rm as}\quad t\rightarrow \infty, 
\end{eqnarray}
where $l_{\alpha}(x)$ is the one-sided stable density with index $\alpha$, which 
depend on $A$ and its Laplace transform is given by
\begin{equation}
\int_0^{\infty} l_{\alpha}(x) e^{-sx} dx = \exp ( -\Gamma (1-\alpha) (As)^{\alpha}).
\end{equation}

In aging renewal processes, the PDF $\psi_0(\tau;t_a)$
%MAYBE CALL THIS FUNC, $\mbox{Dyn}(\tau; t_a)$ 
 of the forward recurrence time, that is the time between
the start of an observation $t=0$ and the first renewal (see Fig. 1), is different from 
(\ref{pdf_interevent_time}). According to \cite{Barkai2003b, God2001}, the double Laplace transform 
of $\psi_0(\tau;t_a)$,
\begin{equation}
\hat{\psi}_0(s,s_a) \equiv \int_0^{\infty} \int_0^{\infty} \psi_0(\tau,t_a) e^{-t_as_a - \tau s} d\tau dt_a,
\end{equation}
is given by
\begin{equation}
\hat{\psi}_0(s,s_a) 
\sim \frac{s_a^{\alpha}-s^{\alpha}}{s_a^{\alpha}(s_a-s)}.
\end{equation}
Furthermore, by Dynkin's limit theorem \cite{Dynkin1961}, the limit PDF $(t_a\rightarrow \infty)$ reads 
\begin{equation}
\psi_0(\tau;t_a) \sim \frac{\sin (\pi \alpha)}{\pi} \frac{t_a^{\alpha}}{\tau^{\alpha} (t_a + \tau)}.
\label{pdf_first_renewal_time}
\end{equation}
The probability of $N_t=0$ is given by $\int_t^{\infty} \psi_0(\tau;t_a)d\tau$, while, for $N_t \geq 1$, 
the probability $\Pr(N_t<n; t_a)$ is represented by the convolution of $\psi_0(\tau;t_a)$ and $\Pr_{t_a=0}(N_{\tau}<n-1)$
\begin{eqnarray}
\Pr(N_t/t^{\alpha}<x; t_a) = 1-m_{\alpha}(T_a)\nonumber\\
+\int_0^{t} \Pr_{t_a=0} (N_{t-\tau} < xt^{\alpha}-1)\psi_0(\tau;t_a)d\tau,
\end{eqnarray}
where $m_{\alpha}(T_a)\equiv \int_0^t \psi_0(\tau;t_a)d\tau$, which is represented by the incomplete beta function,
\begin{equation}
m_{\alpha}(T_a) = \frac{\sin (\pi \alpha)}{\pi} B\left(\frac{1}{T_a+1};1-\alpha,\alpha\right). 
\end{equation}
The $1-m_\alpha(T_a)$ describes trajectories with no renewal events in $(0,t)$,
or in the context of dynamics of maps particles which did not escape
the vicinity of an unstable fixed point in the observation interval.
We note that $T_a$ is the aging ratio, defined in introduction.
In the same way as a calculation of the probability of $N_t/t^{\alpha}$ 
in non-aged renewal processes, we have
\begin{eqnarray}
&&\Pr(N_t/t^{\alpha}<x; t_a) \nonumber\\
&\cong& 1-m_{\alpha}(T_a) + \int_0^{t} \Pr_{t_a=0} \left(\frac{S_n}{n^{1/\alpha}} > 
\frac{t-\tau}{n^{1/\alpha}} \right) \psi_0(\tau;t_a)d\tau \nonumber\\
&\rightarrow& 1-m_{\alpha}(T_a) + \int_0^t \psi_0(\tau;t_a)\int_{\frac{t-\tau}{x^{1/\alpha}t}}^{\infty}  l_{\alpha}(y)dy d\tau,
\label{dist_lim_sigma}
\end{eqnarray}
for $t\rightarrow \infty$. As a result,  
the PDF $p_{\alpha}(\xi; T_a)$ of $\xi\equiv N_t/t^{\alpha}$ in the aging limit, $t_a/t \rightarrow T_a$ 
($t, t_a \rightarrow \infty$), is written as 
%\begin{widetext}
\begin{eqnarray}
p_{\alpha}(\xi;T_a) &=&  \delta (\xi) [1-m_{\alpha}(T_a)] + 
\frac{\sin (\pi \alpha)}{\pi \alpha \xi^{1+1/\alpha}}\nonumber\\
& &\int_0^{1/T_a} \frac{1- T_ay}{y^{\alpha}(1+y)} l_{\alpha}\left(\frac{1-T_ay}{\xi^{1/\alpha}} \right)dy.
\label{PDF_number_renewal}
\end{eqnarray}
%\end{widetext}
%where {\color{red}MAYBE RELATE THIS FUNCTION: $m_{\alpha}(T_a)\equiv \int_0^t \psi_0(\tau;t_a)d\tau$  
%TO DYNKIN BY CALLING IT Dyn?} is the distribution of the first renewal time, 
This result is consistent with \cite{Schulz2012}. 
We note that the distribution depends strongly on $T_a$ even when the total measurement time goes to infinity. 
For $T_a \rightarrow 0$ (non-aging limit), $p_{\alpha}(\xi;T_a)$ converges to the Mittag-Leffler density as expected.

One can show that the mean  $\langle N_t\rangle$ \cite{Schulz2012} is
\begin{equation}
\langle N_t \rangle = \frac{ \{(t+t_a)/A\}^{\alpha }}{\Gamma(\alpha)}
 B\left( \frac{1}{T_a+1}; 1,  \alpha\right).
\end{equation}
It will soon be usefull to define a normalized
variable $\chi = N_t/\langle N_t \rangle$. 
Using Eq.~(\ref{PDF_number_renewal}) and $C_A\equiv t^\alpha/ \langle N_t \rangle$,
we have 
\begin{equation}
P_\alpha (\chi;T_a) = \frac{1}{C_A} p_\alpha (\chi/C_A;T_a).
\label{PDF_normalized}
\end{equation}
We note that PDF $P_\alpha (\chi;T_a)$ has the advantage of being $A$ independent (PDF 
$p_\alpha(\chi;T_a)$ depends on $A$).

\section{Results}
\subsection{Distributional limit theorem in the aging limit}
The distributional limit theorem (\ref{dist_lim_sigma}) in aging renewal processes 
implies that the time average of $\sigma(x_k)$  converges in distribution:
\begin{eqnarray}
\Pr \left( \frac{1}{t^{\alpha}} \sum_{k=0}^{t-1} \sigma(x_k) < x \right) &=&\Pr(N_t/t^{\alpha}<x) \\
&\rightarrow& \int_0^x p_{\alpha}(\xi;T_a)d\xi.
\end{eqnarray}
Because $\sigma(x)$ is an integrable function with respect to an invariant measure,
this distributional limit theorem is a generalization of ADK theorem (\ref{ADK}). 
By the Hopf's ergodic theorem \cite{Hopf1937}, the ratio of 
the sums of arbitrary integrable observation functions $f(x)$ and $\sigma(x)$ converges to a constant
for almost all initial points:
\begin{equation}
\frac{\sum_{k=0}^n f(x_k)}{\sum_{k=0}^n \sigma(x_k)} \rightarrow \frac{\int_0^1 f d\mu}{\int_0^1 \sigma d\mu}
\quad {\rm as}\quad n\rightarrow \infty,
\label{hopf}
\end{equation}
where $\mu$ is an invariant measure. Therefore, we have the following theorem. 
%\begin{theorem}
{\it 
In the aging limit $t_a/t \rightarrow T_a$ as $t_a$ and $t \rightarrow \infty$, 
for all integrable functions $f(x)$ with respect to an invariant measure $\mu$, 
the time average of $f(x)$ converges in distribution: 
\begin{equation}
\Pr \left( \frac{C_fC_A}{t^{\alpha}} \sum_{k=0}^{t-1} f(x_k) < x \right) \rightarrow \int_0^{x} P_{\alpha}(\xi;T_a)d\xi,
\label{dist_lim_th}
\end{equation}
where $C_f= \int_0^1 fd\mu/\mu([\gamma_1,1])$. }
%\end{theorem}
%\begin{remark}
We note that the distribution depends on the aging ratio $T_a$ and $\alpha$ while we do not represent an initial density
explicitly in the left-hand side in (\ref{dist_lim_th}).
%\end{remark}

\subsection{From Dynkin's limit theorem to evolution of density}
Here, we give an explicit representation of an initial density in aging processes. In aging renewal processes, the probability 
that there is no renewal until time $t$ is given by $\int_t^{\infty} \psi_0(\tau;t_a)d\tau$. Corresponding probability in the map 
(\ref{map}) is the probability that trajectories do not escape from the interval $[0,\gamma_1)$, which is given by 
$\int_0^{\gamma_t} \rho(x,t_a)dx$, where $T^t(\gamma_t)=1$ for $\gamma_t<1$. Using a continuous approximation, 
$T'(\gamma_t) \sim \frac{T(\gamma_t) - T(\gamma_{t+1})}{\gamma_t -\gamma_{t+1}}$, near $x \cong 0$ 
and $T(\gamma_t)=\gamma_{t-1}$, we have
\begin{equation}
\frac{\gamma_{t-1}-\gamma_t}{\gamma_t-\gamma_{t+1}} - 1 \sim \left(1+\frac{1}{\alpha}\right) \gamma_t^{1/\alpha}.
\end{equation}
It follows that $\gamma_t \sim \alpha^{\alpha} t^{-\alpha}$  (the rigorous proof is given in \cite{Thaler1983}). 
Therefore, we have the following relation:
\begin{equation}
\int_0^{\alpha^{\alpha}t^{-\alpha}} \rho(x,t_a)dx \sim C' \int_t^{\infty} \psi_0(\tau;t_a)d\tau,
\label{identity}
\end{equation}
where $C'$ is a constant independent of time. % because there exists the cutoff time in the forward recurrence time $\tau$ ($>\alpha$) 
%due to discrete nature of the map. 
 Differentiating both sides of (\ref{identity}) with respect to $t$ and using (\ref{pdf_first_renewal_time}), we have 
\begin{equation}
 \rho( \alpha^{\alpha}/t^{\alpha},t_a) \frac{\alpha^{\alpha+1}}{t^{\alpha+1}}  
\sim C' \frac{\sin(\pi \alpha)}{\pi} \frac{t_a^{\alpha}}{t^{\alpha}(t_a+t)}.
\label{identity1}
\end{equation}
As the result, we obtain an initial density in the aging process:
\begin{equation}
\rho(x,t_a) \sim C \frac{1}{1+t_a x^{1/\alpha}/\alpha}.
\label{ev_density}
\end{equation}
The constant $C$ is the normalization constant which depends on $t_a$, for $t_a \gg 1$
\begin{equation}
C= \left(\int_0^1 \frac{1}{1+t_a x^{1/\alpha}/\alpha}dx\right)^{-1} 
\sim \frac{\sin (\pi \alpha)}{\pi\alpha} \left(\frac{t_a}{\alpha}\right)^{\alpha}.
\end{equation}
Surprisingly, evolutions of the density are in very good agreement with the above estimation even in small numbers of
 iterations and on whole space $[0,1]$ (see Fig. 2%\ref{ev_ID}
 ). The density cannot converge to an invariant density (equilibrium density) 
 and will converge to the delta function $\delta(x)$ as $t_a$ goes to infinity. This is a direct evidence of aging in dynamical 
 systems. We note that the scaled density converges to an infinite invariant density $\tilde{\rho}(x)$ \cite{Thaler2000}:
 \begin{equation}
 \lim_{t_a\rightarrow\infty} t_a^{1-\alpha} \rho (x,t_a) = \tilde{\rho}(x) \sim \frac{\sin (\pi \alpha)}{\pi} \alpha^{-\alpha} x^{-1/\alpha}
  \quad (x\rightarrow 0)
  \label{convergence_invariant_density}
 \end{equation}
 %where {\color{yellow}$c_{\alpha}=( \int_0^{\infty} \frac{dy}{1+y^{1/\alpha}})^{-1}= \frac{\sin (\pi \alpha)}{\pi \alpha}$}.
 Moreover,   by the change of variable, $x =\alpha^{\alpha} y/t_a^{\alpha}$, 
 the scaled density,  $q(y)=\alpha^{\alpha}\rho( \alpha^{\alpha} y/t_a^{\alpha}; t_a)/t_a^{\alpha}$, 
 gives a universal master curve:
\begin{equation}
q(y) = \frac{\sin (\pi \alpha)}{\pi \alpha} \frac{1}{1+y^{1/\alpha}}.
\label{mastercurve}
\end{equation}
This result is consistent with a rigorous result in general intermittent maps by Thaler \cite{Thaler2005}.
Figure 3 %\ref{master_curve} 
shows a convergence of the scaled density to the master curve. 
We note that the master curve is universal in the sense that it does not depends on details of 
the map except for near the fixed point $x=0$.

\begin{figure}
\includegraphics[height=.65\linewidth, angle=0]{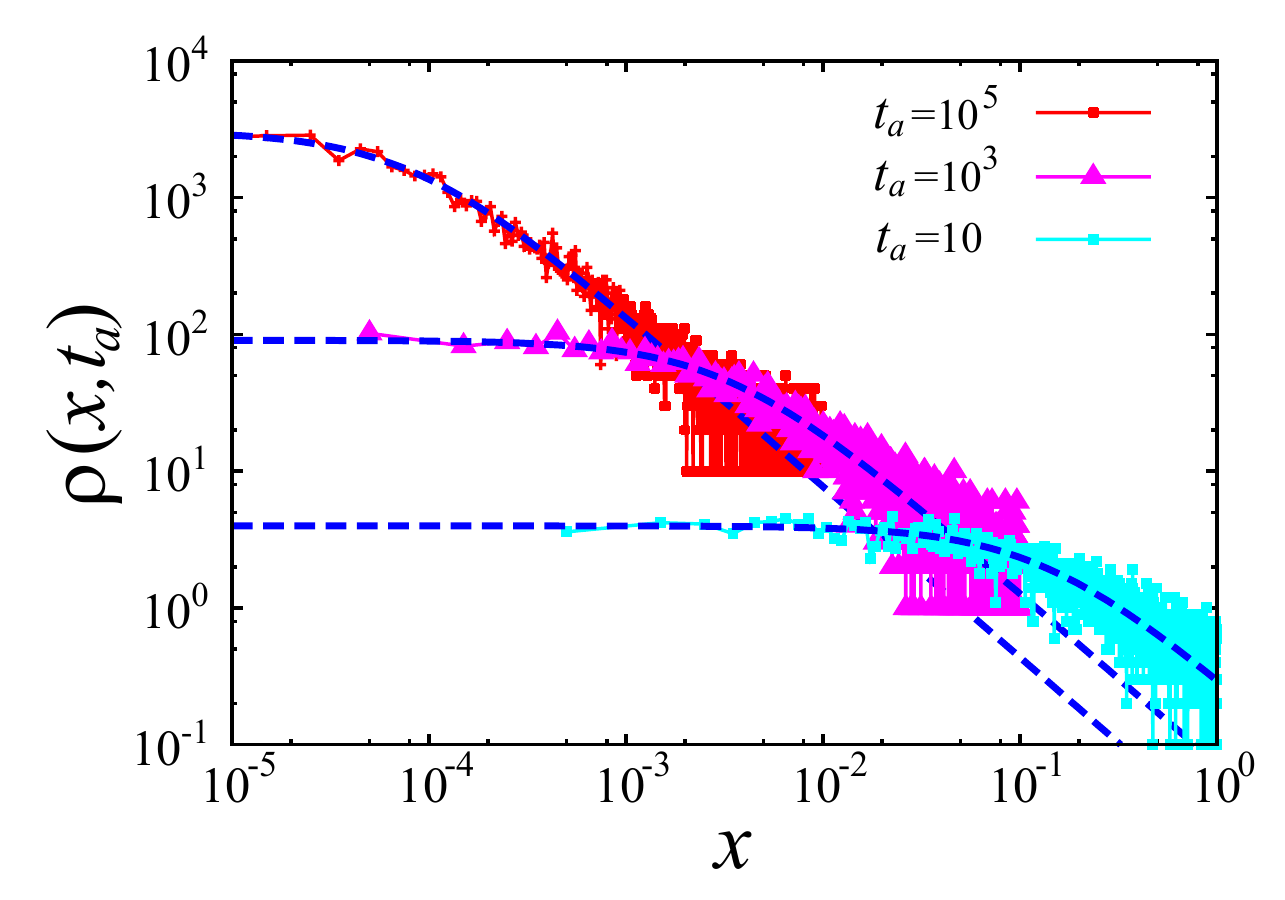}
\caption{Initial densities for different aging times $t_a=10, 10^3$ and $10^5$ ($\alpha=0.8$). Symbols with lines are the results 
of numerical simulations. Blue dashed curves are the theoretical curves without fitting parameters. For all times $t_a$, the densities
are in a nice agreement with the theory.}
\label{ev_ID}
\end{figure}

\begin{figure}
\includegraphics[height=.65\linewidth, angle=0]{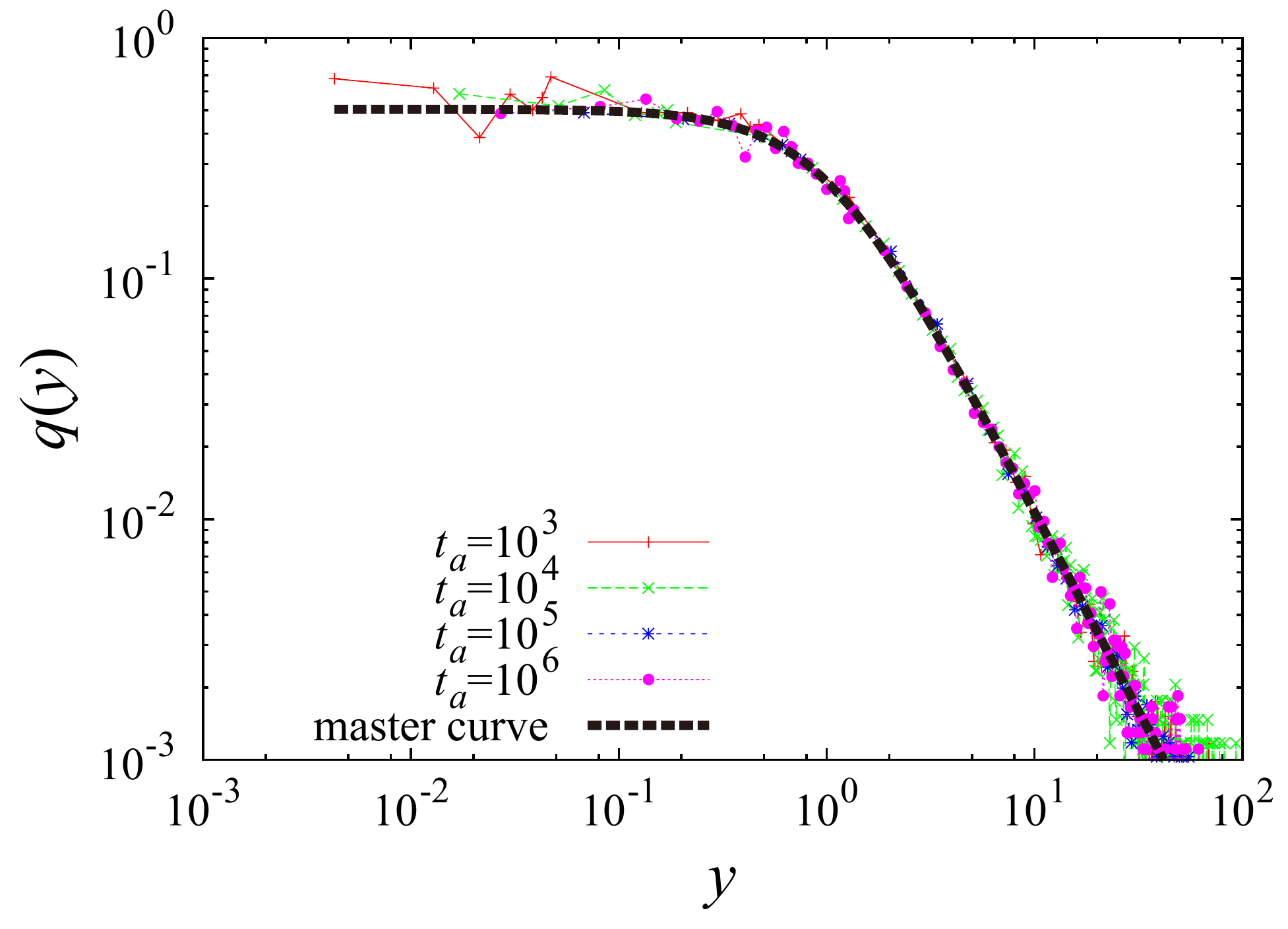}
\caption{Scaled density ($\alpha=0.6$). Dashed line is the master curve Eq.~(\ref{mastercurve}) for $\alpha=0.6$. 
The scaled density approach to the master curve as $t_a\rightarrow \infty$.
%I SEE A LOT OF NOISE, REFEREES CAN GET UPSET.... ALSO I SUGGEST NOT TO USE $\alpha=0.8$ 
%SINCE CONVERGENCE BECOMES SLOW WHEN $\alpha \to 1$, BETTER USE $\alpha =0.6$.
}
\label{master_curve}
\end{figure}

\subsection{Dynamical instability}
\begin{figure}
\includegraphics[height=.65\linewidth, angle=0]{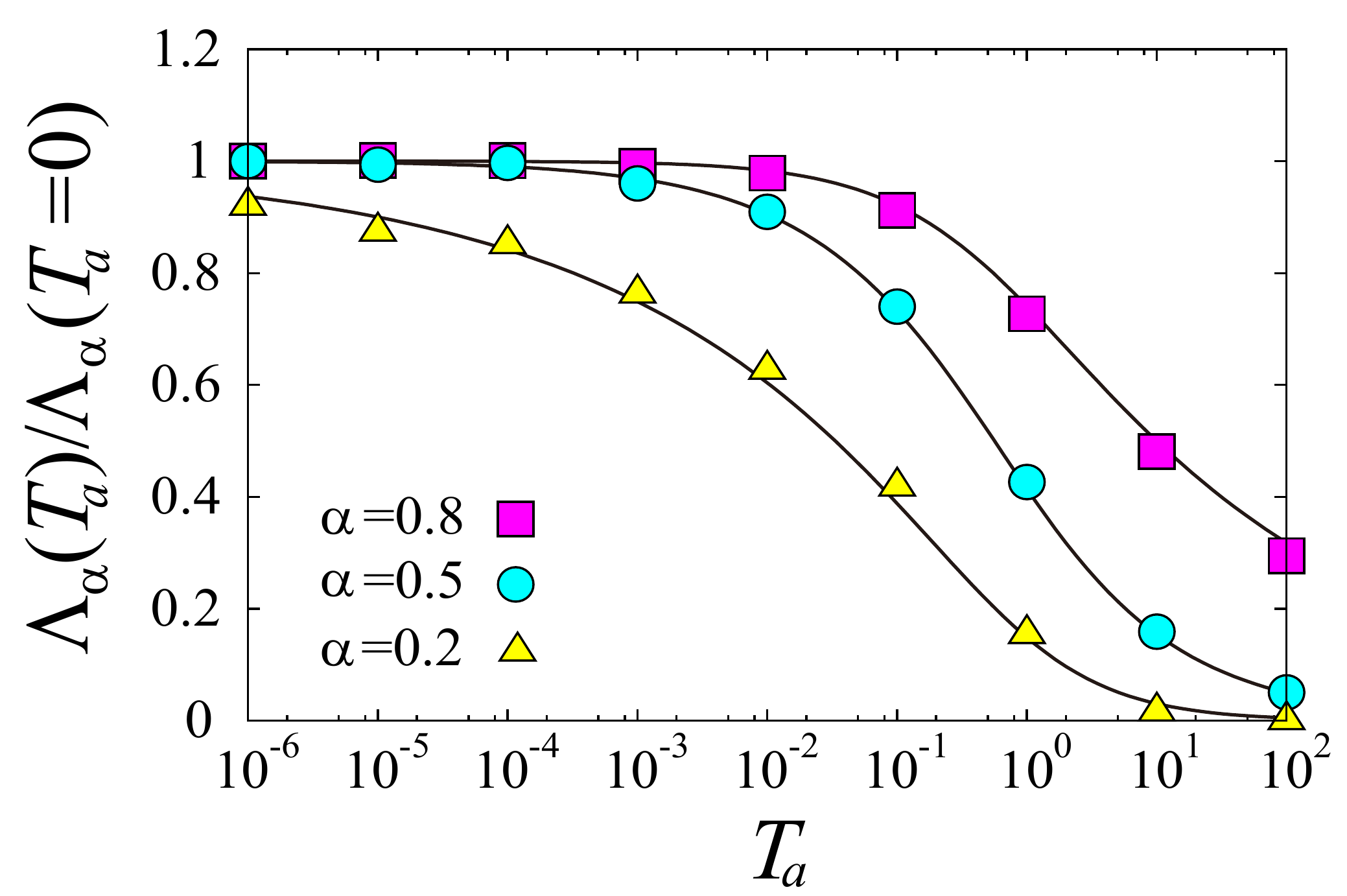}
\caption{Generalized Lyapunov exponent as a function of the aging ratio $T_a$  ($\alpha=0.8$, $0.5$, and $0.2$).
Different symbols are the results of numerical simulations for different $\alpha$, where the total measurement time 
$t$ is fixed as $10^6$. 
Curves are theoretical ones without fitting parameter 
($\Lambda_\alpha(T_a=0)$ is obtained numerically).
%I SUGGEST TO RENAME TO $\langle \Lambda_{\alpha} (T_a=0) \rangle$,
%{\color{blue} MORE IMPOTANTLY I DO NOT SEE WHY YOU CALL THE PLOTTED FUNCTION A FIT. 
%{\bf your reply:}  Because we do not have $\Lambda(T_a=0)$ numerically, $\Lambda(T_a=0)$ is a fitting parameter. 
%(my reply to this: SO WHY NOT GET IT FROM SIMULATIONS? )} 
Aging strongly affects the dynamical instability 
when the aging time $t_a$ is larger than the measurement time $t$.}
\label{GLYP}
\end{figure}

To investigate an effect of the aging on the dynamical instability, we consider the Lyapunov exponent.
In general, weakly chaotic systems with infinite invariant measures 
have  zero Lyapunov exponent
\cite{Gaspard1988, Akimoto2010a, Ignaccolo2001, Korabel2009}. 
However, these dynamical instabilities are known as a subexponential instability quantified by the 
generalized Lyapunov exponent 
$\Lambda_{\alpha}$ \cite{Akimoto2010a, Korabel2009}, which is defined as the average of the normalized 
Lyapunov exponent, $\Lambda_{\alpha}\equiv \langle \lambda_{\alpha}\rangle$, where
 $\langle \cdot \rangle$ is an average with respect to an initial density and 
\begin{equation}
\lambda_{\alpha} %\equiv \lim_{t\rightarrow \infty, t_a=0}  \lambda_{\alpha}(t;t_a) 
\equiv \lim_{t\rightarrow \infty} 
\frac{1}{t^{\alpha}} \sum_{k=0}^{t-1} \ln |T'(x_k)|, \quad (t_a=0).
\end{equation}
In non-aged systems, $\Lambda_{\alpha}$ does not depend on an initial density with the aid of ADK theorem.
To investigate an effect of aging, we consider the generalized Lyapunov exponent in the aging limit 
$t_a/t \rightarrow T_a$,  
$\Lambda_{\alpha}(T_a) \equiv  \langle \lambda_{\alpha}(T_a) \rangle$, where 
\begin{equation}
 \lambda_{\alpha}(T_a) \equiv  \lim_{t\rightarrow \infty} 
\frac{1}{t^{\alpha}} \sum_{k=0}^{t-1} \ln |T'(x_k)|, \quad (t_a/t \rightarrow T_a).
\end{equation}
%where $\langle \cdot \rangle_{t_a}$ is an average with respect to an initial density with an aging time $t_a$. 
The aging generalized Lyapunov exponent is represented as
\begin{equation}
\Lambda_{\alpha}(T_a) \cong \frac{1}{t^{\alpha}} \int_0^t  \int_0^1 g(x) \rho(x,t'+t_a) dx dt',
\end{equation}
where $g(x) = \ln |T'(x)|$.
 Using the limit density (\ref{convergence_invariant_density}), we obtain
\begin{equation}
\Lambda_{\alpha}(T_a) 
\cong \frac{(t+t_a)^{\alpha}-t_a^{\alpha}}{\alpha t^{\alpha}}\int_0^1 g(x) \tilde{\rho}(x) dx .
\end{equation}
for $t\gg 1$ and $t_a \gg 1$ with $T_a$ fixed. Here $\tilde{\rho}(x) = \lim_{t_a\to \infty} t_a^{1-\alpha} \rho (x,0) $
defines a unique  infinite invariant density. 
%which would be related to {\it pointwise dual ergodicity} \cite{Aaronson1997}.
According to ADK theorem, 
the non-aging generalized Lyapunov exponent ($t_a=0$) is given by 
\cite{Korabel2009}
\begin{equation}
\Lambda_{\alpha} (T_a=0)= \frac{1}{\alpha} \int_0^1g(x) \tilde{\rho}(x)dx. 
\end{equation} 
%
%THE FOLLOWING SENTENCE IS UNCLEAR. 
%{\color{red} We note that the invariant density $\tilde{\rho}(x)$ is determined up to a multiplicative 
%constant whereas $a\tilde{\rho}(x)$ for a constant $a>0$ is also an infinite invariant density. }
%I SUGGEST
 In the aging limit, this relation
is generalized as
\begin{equation}
\Lambda_{\alpha}(T_a) = \frac{(1+T_a)^{\alpha} -T_a^{\alpha}}{\alpha} \int_0^1 g(x) \tilde{\rho}(x)dx. 
%\Lambda_{\alpha} (T_a=0).
\label{aging_LYP}
\end{equation}
%This is also derived by the Hopf's ergodic theorem {\bf\ref{hopf}} and the first moment of $N_t$. 
This aging effect on the generalized Lyapunov exponent has been confirmed numerically (Fig.~\ref{GLYP}).
The result means that the dynamical instability becomes weak as the system ages. 

%NOW THAT WE HAVE THE AVERAGE OF $\langle \Lambda_\alpha (T_a) \rangle$ EQ.
%35 IT IS BETTER TO DEFINE A NEW VARAIBLE WITH MEAN EQUAL 1,
%$\xi = \lambda_\alpha / \langle \lambda_\alpha \rangle$ and GIVE ITS DISTRIBUTION OR PDF. 
Using 
the distributional limit theorem (\ref{dist_lim_th}), we obtain the PDF of the normalized Lyapunov exponent 
$\chi = \lambda_\alpha(T_a) /  \Lambda_\alpha (T_a) $, given by PDF (\ref{PDF_normalized}), $P_{\alpha}(\chi; T_a)$.
%
%\begin{equation}
%P_{\alpha}(\chi) = \frac{\Lambda_\alpha (T_a)}{C_g}p_{\alpha}(\Lambda_\alpha (T_a) \chi/C_g;T_a). 
%\end{equation}}
%
%THIS IS UNCLEAR SINCE $C_g$ IS UNKNOWN. INSTEAD IN DISTRIBUTION
%$\chi = N_t / \langle N_t \rangle$ SO ITS PDF IS GIVEN BY MY SUGGESTED
%EQ. (please verify) 
%(\label{eqNEWELI}). THE ADVANTAGE IS THAT NOW THERE ARE NO 
%FREE PARAMETERS LIKE $A$ AND $C_g$. WHAT DO YOU THINK?
As shown in Fig.~5%\ref{pdf_lyapunov}
, the PDF does not depend on the total measurement time $t$ if the aging ratio 
$T_a$ is fixed, and 
the strength of the delta peak at $\chi=0$ 
is increased as  the aging ratio $T_a$ is made larger. 
This delta peak corresponds to trajectories that do not escape from $[0,\gamma_1]$ until time $t$, which
 weakens the dynamical instability.  These trajectories are regular rather than chaotic (not even weakly chaotic). 
In fact, these trajectories can be treated by a continuous approximation. For $x_t \cong 0$, intermittent maps we consider
  can be written as an ordinary differential equation \cite{ Manneville1980, Geisel1984}:
\begin{equation}
\frac{dx_t}{dt} = a_0 x_t^{1+1/\alpha} \quad (x_t \cong 0).
\end{equation}
The solution is given by
%THIS SOLUTION WAS DISCUSSED BY GRIGOLINI AND COMPANY SEE MY PRE WITH KORABEL FOR REF. 
\begin{equation}
x_t = x_0 (1- \alpha^{-1}x_0^{1/\alpha} a_0 t)^{-\alpha}
\end{equation}
for $x_t \ll 1$ or $t\ll  \alpha x_0^{-1/\alpha}/a_0$. The difference of nearby trajectories $x_t$ and $x'_t$ such that $x_0$ and 
$x'_0=x_0 + \delta x$ ($\delta x \ll x_0$) is described as 
\begin{equation}
|x_t - x'_t| \cong \delta x (1 + x_0^{1/\alpha} (1+\alpha^{-1}) a_0 t).
\end{equation}
In the aging limit, these regular motions appear even when time $t$ goes to infinity. Therefore, in the aging dynamical systems, 
regular motions (RMs) and weak chaotic motions (WCMs) intrinsically coexist in the aging limit. 
This implies that when $t_a>>t$ a large fraction of particles
are located close to the unstable fixed point and hence their
motion is regular.

\newpage

\begin{figure}
\centerline{\includegraphics[width=.45\textwidth, angle=0]{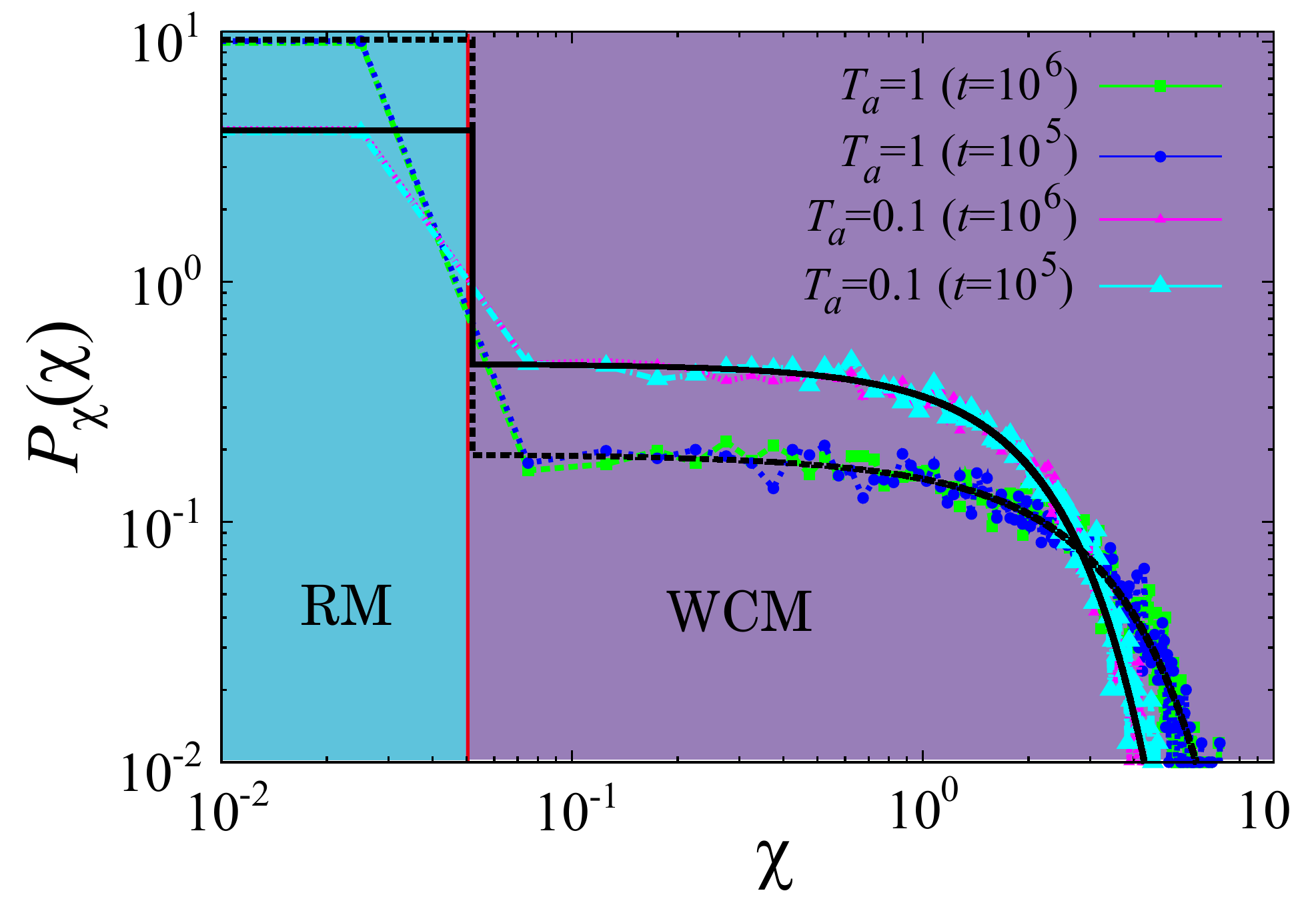}}
\caption{Probability density function $P_{\chi}({\chi})$ of the normalized Lyapunov exponent  ($T_a=0.1$ 
and $1$, $t=10^5$ and $10^6$, and $\alpha=0.5$). 
Symbols with lines are the results of numerical simulations. Solid curves are the theoretical ones calculated by a numerical
integration of the convolution in (\ref{PDF_number_renewal}) using the stable density with index $1/2$, i.e., 
$l_{1/2}(x) = c\exp(-c^2/(2x))/\sqrt{2 \pi x^3}$, where $c$ is a scaling parameter depending on $T_a$. 
Probability of regular motions is increased when the system ages. Almost all trajectories in the left-hand side of the red line exhibit
regular motions (RMs) while trajectories in the right-hand side are weakly chaotic motions (WCMs), where the value of the red line 
($\chi=0.05$)  represents the bin size of the PDF with histogram (the bin is needed to graphically
 represent a delta function).
Theory is in good agreement with the numerical results including the delta distribution.}
\label{pdf_lyapunov}
\end{figure}

\section{Discussion}
We have shown the distribution of time averages of integrable functions in the aging limit, which is a generalization of Aaronson's 
Darling-Kac theorem. Although we use intermittent maps, 
this generalization will be valid for all weakly chaotic maps with 
infinite invariant measures (more precisely, a conservative, ergodic, measure preserving transformation)
because aging in dynamical systems implies that the density does not converges to an equilibrium density  even when time goes to infinity 
(non-equilibrium non-stationary density). 
%{\color{blue}THE FOLLOWING STATEMENT IS
%UNCLEAR, THE BEHAVIOT IS NOT SO GENERAL, IT IS LIMITED.
% Such non-equilibrium non-stationary {\color{red}densities} appear generally in 
%a conservative, ergodic, measure preserving transformation. 
In fact, we have confirmed the distributional aging limit theorem in the Boole transformation \cite{Aaronson1997}.  
%Although we focus on infinite measure systems, an apparent aging appears in an intermittent map with $1<\alpha<2$ because a convergence to a equilibrium density is extremely slow. 

\section{Conclusion}
The aging ratio $T_a$ plays an important role in characterizing aging systems. In the aging limit,
the distribution of time-averaged observables converges to a universal distribution, which is determined by the aging ratio $T_a$ and 
the exponent $\alpha$ characterizing  infinite invariant measures in dynamical systems. 
The mathematical
basis of this  universal distribution is both the generalized 
 central limit theorem and Dynkin's theorem for the forward recurrence
time. We have also shown how to use
 the infinite invariant density 
to calculate statistical averages like the measure of
separation $\Lambda_\alpha$ within the aging regime.
From Eq.~(\ref{aging_LYP}) we see   an aged dependent
pre-factor  $(1+T_a)^\alpha - (T_a)^\alpha$ which multiplies the 
non-aged average. 
Similar averages hold for other time-averaged observables,
which are integrable with respect to the infinite invariant density. 
Thus the infinite invariant density plays an important role 
for determination of ergodic properties of
aging processes. 

We have found that the dynamical instability is clearly divided into two different 
instabilities, i.e., regular motions and weakly chaotic motions, in the aging limit.
Coexistence of regular and chaotic motions is reminiscent of generic Hamiltonian 
systems. However, the meaning of the coexistence is completely different. 
In aging dynamical systems, the probability of finding regular motions is increased according to 
the aging ratio $T_a$, 
whereas regular and chaotic phase spaces do not depend on $T_a$ in generic Hamiltonian systems.

\begin{acknowledgments}
This work was partially supported by 
Grant-in-Aid for Young Scientists (B) No. 22740262 (to T.A.).
\end{acknowledgments}

%section {bibliography}

%\bibliography{akimoto}
%\begin{thebibliography}{24}%

%\end{thebibliography}%

\bibliography{aging}

%merlin.mbs apsrev4-1.bst 2010-07-25 4.21a (PWD, AO, DPC) hacked
%Control: key (0)
%Control: author (8) initials jnrlst
%Control: editor formatted (1) identically to author
%Control: production of article title (-1) disabled
%Control: page (0) single
%Control: year (1) truncated
%Control: production of eprint (0) enabled
\begin{thebibliography}{37}%
\makeatletter
\providecommand \@ifxundefined [1]{%
 \@ifx{#1\undefined}
}%
\providecommand \@ifnum [1]{%
 \ifnum #1\expandafter \@firstoftwo
 \else \expandafter \@secondoftwo
 \fi
}%
\providecommand \@ifx [1]{%
 \ifx #1\expandafter \@firstoftwo
 \else \expandafter \@secondoftwo
 \fi
}%
\providecommand \natexlab [1]{#1}%
\providecommand \enquote  [1]{``#1''}%
\providecommand \bibnamefont  [1]{#1}%
\providecommand \bibfnamefont [1]{#1}%
\providecommand \citenamefont [1]{#1}%
\providecommand \href@noop [0]{\@secondoftwo}%
\providecommand \href [0]{\begingroup \@sanitize@url \@href}%
\providecommand \@href[1]{\@@startlink{#1}\@@href}%
\providecommand \@@href[1]{\endgroup#1\@@endlink}%
\providecommand \@sanitize@url [0]{\catcode `\\12\catcode `\$12\catcode
  `\&12\catcode `\#12\catcode `\^12\catcode `\_12\catcode `\%12\relax}%
\providecommand \@@startlink[1]{}%
\providecommand \@@endlink[0]{}%
\providecommand \url  [0]{\begingroup\@sanitize@url \@url }%
\providecommand \@url [1]{\endgroup\@href {#1}{\urlprefix }}%
\providecommand \urlprefix  [0]{URL }%
\providecommand \Eprint [0]{\href }%
\providecommand \doibase [0]{http://dx.doi.org/}%
\providecommand \selectlanguage [0]{\@gobble}%
\providecommand \bibinfo  [0]{\@secondoftwo}%
\providecommand \bibfield  [0]{\@secondoftwo}%
\providecommand \translation [1]{[#1]}%
\providecommand \BibitemOpen [0]{}%
\providecommand \bibitemStop [0]{}%
\providecommand \bibitemNoStop [0]{.\EOS\space}%
\providecommand \EOS [0]{\spacefactor3000\relax}%
\providecommand \BibitemShut  [1]{\csname bibitem#1\endcsname}%
\let\auto@bib@innerbib\@empty
%</preamble>
\bibitem [{\citenamefont {Bouchaud}(1992)}]{Bouchaud1992}%
  \BibitemOpen
  \bibfield  {author} {\bibinfo {author} {\bibfnamefont {J.}~\bibnamefont
  {Bouchaud}},\ }\href@noop {} {\bibfield  {journal} {\bibinfo  {journal} {J.
  Phys. I (France)}\ }\textbf {\bibinfo {volume} {2}},\ \bibinfo {pages} {1705}
  (\bibinfo {year} {1992})}\BibitemShut {NoStop}%
\bibitem [{\citenamefont {Takeuchi}\ and\ \citenamefont
  {Sano}(2012)}]{Takeuchi2012}%
  \BibitemOpen
  \bibfield  {author} {\bibinfo {author} {\bibfnamefont {K.}~\bibnamefont
  {Takeuchi}}\ and\ \bibinfo {author} {\bibfnamefont {M.}~\bibnamefont
  {Sano}},\ }\href@noop {} {\bibfield  {journal} {\bibinfo  {journal} {J. Stat.
  Phys.}\ ,\ \bibinfo {pages} {1}} (\bibinfo {year} {2012})}\BibitemShut
  {NoStop}%
\bibitem [{\citenamefont {Brokmann}\ \emph {et~al.}(2003)\citenamefont
  {Brokmann}, \citenamefont {Hermier}, \citenamefont {Messin}, \citenamefont
  {Desbiolles}, \citenamefont {Bouchaud},\ and\ \citenamefont
  {Dahan}}]{Brok2003}%
  \BibitemOpen
  \bibfield  {author} {\bibinfo {author} {\bibfnamefont {X.}~\bibnamefont
  {Brokmann}}, \bibinfo {author} {\bibfnamefont {J.-P.}\ \bibnamefont
  {Hermier}}, \bibinfo {author} {\bibfnamefont {G.}~\bibnamefont {Messin}},
  \bibinfo {author} {\bibfnamefont {P.}~\bibnamefont {Desbiolles}}, \bibinfo
  {author} {\bibfnamefont {J.-P.}\ \bibnamefont {Bouchaud}}, \ and\ \bibinfo
  {author} {\bibfnamefont {M.}~\bibnamefont {Dahan}},\ }\href@noop {}
  {\bibfield  {journal} {\bibinfo  {journal} {Phys. Rev. Lett.}\ }\textbf
  {\bibinfo {volume} {90}},\ \bibinfo {pages} {120601} (\bibinfo {year}
  {2003})}\BibitemShut {NoStop}%
\bibitem [{\citenamefont {Margolin}\ and\ \citenamefont
  {Barkai}(2004)}]{Margolin2004}%
  \BibitemOpen
  \bibfield  {author} {\bibinfo {author} {\bibfnamefont {G.}~\bibnamefont
  {Margolin}}\ and\ \bibinfo {author} {\bibfnamefont {E.}~\bibnamefont
  {Barkai}},\ }\href@noop {} {\bibfield  {journal} {\bibinfo  {journal} {J.
  Chem. Phys.}\ }\textbf {\bibinfo {volume} {121}},\ \bibinfo {pages} {1566}
  (\bibinfo {year} {2004})}\BibitemShut {NoStop}%
\bibitem [{\citenamefont {Weigel}\ \emph {et~al.}(2011)\citenamefont {Weigel},
  \citenamefont {Simon}, \citenamefont {Tamkun},\ and\ \citenamefont
  {Krapf}}]{Weigel2011}%
  \BibitemOpen
  \bibfield  {author} {\bibinfo {author} {\bibfnamefont {A.}~\bibnamefont
  {Weigel}}, \bibinfo {author} {\bibfnamefont {B.}~\bibnamefont {Simon}},
  \bibinfo {author} {\bibfnamefont {M.}~\bibnamefont {Tamkun}}, \ and\ \bibinfo
  {author} {\bibfnamefont {D.}~\bibnamefont {Krapf}},\ }\href@noop {}
  {\bibfield  {journal} {\bibinfo  {journal} {Proc. Natl. Acad. Sci. USA}\
  }\textbf {\bibinfo {volume} {108}},\ \bibinfo {pages} {6438} (\bibinfo {year}
  {2011})}\BibitemShut {NoStop}%
\bibitem [{\citenamefont {Burov}\ \emph {et~al.}(2010)\citenamefont {Burov},
  \citenamefont {Metzler},\ and\ \citenamefont {Barkai}}]{Burov2010}%
  \BibitemOpen
  \bibfield  {author} {\bibinfo {author} {\bibfnamefont {S.}~\bibnamefont
  {Burov}}, \bibinfo {author} {\bibfnamefont {R.}~\bibnamefont {Metzler}}, \
  and\ \bibinfo {author} {\bibfnamefont {E.}~\bibnamefont {Barkai}},\
  }\href@noop {} {\bibfield  {journal} {\bibinfo  {journal} {Proc. Nat. Acad.
  Sci. USA}\ }\textbf {\bibinfo {volume} {107}},\ \bibinfo {pages} {13228}
  (\bibinfo {year} {2010})}\BibitemShut {NoStop}%
\bibitem [{\citenamefont {Schulz}\ \emph {et~al.}(2012)\citenamefont {Schulz},
  \citenamefont {Barkai},\ and\ \citenamefont {Metzler}}]{Schulz2012}%
  \BibitemOpen
  \bibfield  {author} {\bibinfo {author} {\bibfnamefont {J.}~\bibnamefont
  {Schulz}}, \bibinfo {author} {\bibfnamefont {E.}~\bibnamefont {Barkai}}, \
  and\ \bibinfo {author} {\bibfnamefont {R.}~\bibnamefont {Metzler}},\
  }\href@noop {} {\bibfield  {journal} {\bibinfo  {journal} {arXiv:1204.0878}\
  } (\bibinfo {year} {2012})}\BibitemShut {NoStop}%
\bibitem [{\citenamefont {Aaronson}(1997)}]{Aaronson1997}%
  \BibitemOpen
  \bibfield  {author} {\bibinfo {author} {\bibfnamefont {J.}~\bibnamefont
  {Aaronson}},\ }\href@noop {} {\emph {\bibinfo {title} {An Introduction to
  Infinite Ergodic Theory}}}\ (\bibinfo  {publisher} {American Mathematical
  Society},\ \bibinfo {address} {Province},\ \bibinfo {year}
  {1997})\BibitemShut {NoStop}%
\bibitem [{\citenamefont {Aizawa}\ \emph {et~al.}(1989)\citenamefont {Aizawa},
  \citenamefont {Kikuchi}, \citenamefont {Harayama}, \citenamefont {Yamamoto},
  \citenamefont {Ota},\ and\ \citenamefont {Tanaka}}]{Aizawa1989a}%
  \BibitemOpen
  \bibfield  {author} {\bibinfo {author} {\bibfnamefont {Y.}~\bibnamefont
  {Aizawa}}, \bibinfo {author} {\bibfnamefont {Y.}~\bibnamefont {Kikuchi}},
  \bibinfo {author} {\bibfnamefont {T.}~\bibnamefont {Harayama}}, \bibinfo
  {author} {\bibfnamefont {K.}~\bibnamefont {Yamamoto}}, \bibinfo {author}
  {\bibfnamefont {M.}~\bibnamefont {Ota}}, \ and\ \bibinfo {author}
  {\bibfnamefont {K.}~\bibnamefont {Tanaka}},\ }\href@noop {} {\bibfield
  {journal} {\bibinfo  {journal} {Prog. Theor. Phys. Suppl.}\ }\textbf
  {\bibinfo {volume} {98}},\ \bibinfo {pages} {36} (\bibinfo {year}
  {1989})}\BibitemShut {NoStop}%
\bibitem [{\citenamefont {Gaspard}\ and\ \citenamefont
  {Wang}(1988)}]{Gaspard1988}%
  \BibitemOpen
  \bibfield  {author} {\bibinfo {author} {\bibfnamefont {P.}~\bibnamefont
  {Gaspard}}\ and\ \bibinfo {author} {\bibfnamefont {X.~J.}\ \bibnamefont
  {Wang}},\ }\href@noop {} {\bibfield  {journal} {\bibinfo  {journal} {Proc.
  Natl. Acad. Sci. USA}\ }\textbf {\bibinfo {volume} {85}},\ \bibinfo {pages}
  {4591} (\bibinfo {year} {1988})}\BibitemShut {NoStop}%
\bibitem [{\citenamefont {Akimoto}\ and\ \citenamefont
  {Aizawa}(2010)}]{Akimoto2010a}%
  \BibitemOpen
  \bibfield  {author} {\bibinfo {author} {\bibfnamefont {T.}~\bibnamefont
  {Akimoto}}\ and\ \bibinfo {author} {\bibfnamefont {Y.}~\bibnamefont
  {Aizawa}},\ }\href@noop {} {\bibfield  {journal} {\bibinfo  {journal}
  {Chaos}\ }\textbf {\bibinfo {volume} {20}},\ \bibinfo {pages} {033110}
  (\bibinfo {year} {2010})}\BibitemShut {NoStop}%
\bibitem [{\citenamefont {Barkai}(2003)}]{Barkai2003}%
  \BibitemOpen
  \bibfield  {author} {\bibinfo {author} {\bibfnamefont {E.}~\bibnamefont
  {Barkai}},\ }\href@noop {} {\bibfield  {journal} {\bibinfo  {journal} {Phys.
  Rev. Lett.}\ }\textbf {\bibinfo {volume} {90}},\ \bibinfo {pages} {104101}
  (\bibinfo {year} {2003})}\BibitemShut {NoStop}%
\bibitem [{\citenamefont {Birkhoff}(1931)}]{Birkhoff1931}%
  \BibitemOpen
  \bibfield  {author} {\bibinfo {author} {\bibfnamefont {G.~D.}\ \bibnamefont
  {Birkhoff}},\ }\href@noop {} {\bibfield  {journal} {\bibinfo  {journal}
  {Proc. Natl. Acad. Sci. USA}\ }\textbf {\bibinfo {volume} {17}},\ \bibinfo
  {pages} {656} (\bibinfo {year} {1931})}\BibitemShut {NoStop}%
\bibitem [{\citenamefont {Aaronson}(1981)}]{Aaronson1981}%
  \BibitemOpen
  \bibfield  {author} {\bibinfo {author} {\bibfnamefont {J.}~\bibnamefont
  {Aaronson}},\ }\href@noop {} {\bibfield  {journal} {\bibinfo  {journal} {J.
  D'Analyse Math.}\ }\textbf {\bibinfo {volume} {39}},\ \bibinfo {pages} {203}
  (\bibinfo {year} {1981})}\BibitemShut {NoStop}%
\bibitem [{\citenamefont {Darling}\ and\ \citenamefont
  {Kac}(1957)}]{Darling1957}%
  \BibitemOpen
  \bibfield  {author} {\bibinfo {author} {\bibfnamefont {D.~A.}\ \bibnamefont
  {Darling}}\ and\ \bibinfo {author} {\bibfnamefont {M.}~\bibnamefont {Kac}},\
  }\href@noop {} {\bibfield  {journal} {\bibinfo  {journal} {Trans. Am. Math.
  Soc.}\ }\textbf {\bibinfo {volume} {84}},\ \bibinfo {pages} {444} (\bibinfo
  {year} {1957})}\BibitemShut {NoStop}%
\bibitem [{\citenamefont {Akimoto}(2008)}]{Akimoto2008}%
  \BibitemOpen
  \bibfield  {author} {\bibinfo {author} {\bibfnamefont {T.}~\bibnamefont
  {Akimoto}},\ }\href@noop {} {\bibfield  {journal} {\bibinfo  {journal} {J.
  Stat. Phys.}\ }\textbf {\bibinfo {volume} {132}},\ \bibinfo {pages} {171}
  (\bibinfo {year} {2008})}\BibitemShut {NoStop}%
\bibitem [{\citenamefont {Akimoto}\ and\ \citenamefont
  {Miyaguchi}(2010)}]{Akimoto2010}%
  \BibitemOpen
  \bibfield  {author} {\bibinfo {author} {\bibfnamefont {T.}~\bibnamefont
  {Akimoto}}\ and\ \bibinfo {author} {\bibfnamefont {T.}~\bibnamefont
  {Miyaguchi}},\ }\href@noop {} {\bibfield  {journal} {\bibinfo  {journal}
  {Phys. Rev. E}\ }\textbf {\bibinfo {volume} {82}},\ \bibinfo {pages}
  {030102(R)} (\bibinfo {year} {2010})}\BibitemShut {NoStop}%
\bibitem [{\citenamefont {Akimoto}(2012)}]{Akimoto2012}%
  \BibitemOpen
  \bibfield  {author} {\bibinfo {author} {\bibfnamefont {T.}~\bibnamefont
  {Akimoto}},\ }\href {\doibase 10.1103/PhysRevLett.108.164101} {\bibfield
  {journal} {\bibinfo  {journal} {Phys. Rev. Lett.}\ }\textbf {\bibinfo
  {volume} {108}},\ \bibinfo {pages} {164101} (\bibinfo {year}
  {2012})}\BibitemShut {NoStop}%
\bibitem [{\citenamefont {Thaler}(1980)}]{Thaler1980}%
  \BibitemOpen
  \bibfield  {author} {\bibinfo {author} {\bibfnamefont {M.}~\bibnamefont
  {Thaler}},\ }\href@noop {} {\bibfield  {journal} {\bibinfo  {journal} {Israel
  Journal of Mathematics}\ }\textbf {\bibinfo {volume} {37}},\ \bibinfo {pages}
  {303} (\bibinfo {year} {1980})}\BibitemShut {NoStop}%
\bibitem [{\citenamefont {Pomeau}\ and\ \citenamefont
  {Manneville}(1980)}]{Pomeau1980}%
  \BibitemOpen
  \bibfield  {author} {\bibinfo {author} {\bibfnamefont {Y.}~\bibnamefont
  {Pomeau}}\ and\ \bibinfo {author} {\bibfnamefont {P.}~\bibnamefont
  {Manneville}},\ }\href@noop {} {\bibfield  {journal} {\bibinfo  {journal}
  {Communications in Mathematical Physics}\ }\textbf {\bibinfo {volume} {74}},\
  \bibinfo {pages} {189} (\bibinfo {year} {1980})}\BibitemShut {NoStop}%
\bibitem [{\citenamefont {Manneville}(1980)}]{Manneville1980}%
  \BibitemOpen
  \bibfield  {author} {\bibinfo {author} {\bibfnamefont {P.}~\bibnamefont
  {Manneville}},\ }\href@noop {} {\bibfield  {journal} {\bibinfo  {journal} {J.
  Phys. (Paris)}\ }\textbf {\bibinfo {volume} {41}},\ \bibinfo {pages} {1235}
  (\bibinfo {year} {1980})}\BibitemShut {NoStop}%
\bibitem [{\citenamefont {Aizawa}(1984)}]{Aizawa1984}%
  \BibitemOpen
  \bibfield  {author} {\bibinfo {author} {\bibfnamefont {Y.}~\bibnamefont
  {Aizawa}},\ }\href@noop {} {\bibfield  {journal} {\bibinfo  {journal} {Prog.
  Theor. Phys.}\ }\textbf {\bibinfo {volume} {72}},\ \bibinfo {pages} {659}
  (\bibinfo {year} {1984})}\BibitemShut {NoStop}%
\bibitem [{\citenamefont {Geisel}\ and\ \citenamefont
  {Thomae}(1984)}]{Geisel1984}%
  \BibitemOpen
  \bibfield  {author} {\bibinfo {author} {\bibfnamefont {T.}~\bibnamefont
  {Geisel}}\ and\ \bibinfo {author} {\bibfnamefont {S.}~\bibnamefont
  {Thomae}},\ }\href@noop {} {\bibfield  {journal} {\bibinfo  {journal} {Phys.
  Rev. Lett.}\ }\textbf {\bibinfo {volume} {52}},\ \bibinfo {pages} {1936}
  (\bibinfo {year} {1984})}\BibitemShut {NoStop}%
\bibitem [{\citenamefont {Geisel}\ \emph {et~al.}(1985)\citenamefont {Geisel},
  \citenamefont {Nierwetberg},\ and\ \citenamefont {Zacherl}}]{Geisel1985}%
  \BibitemOpen
  \bibfield  {author} {\bibinfo {author} {\bibfnamefont {T.}~\bibnamefont
  {Geisel}}, \bibinfo {author} {\bibfnamefont {J.}~\bibnamefont {Nierwetberg}},
  \ and\ \bibinfo {author} {\bibfnamefont {A.}~\bibnamefont {Zacherl}},\ }\href
  {\doibase 10.1103/PhysRevLett.54.616} {\bibfield  {journal} {\bibinfo
  {journal} {Phys. Rev. Lett.}\ }\textbf {\bibinfo {volume} {54}},\ \bibinfo
  {pages} {616} (\bibinfo {year} {1985})}\BibitemShut {NoStop}%
\bibitem [{\citenamefont {Zumofen}\ and\ \citenamefont
  {Klafter}(1993)}]{Zumofen1993}%
  \BibitemOpen
  \bibfield  {author} {\bibinfo {author} {\bibfnamefont {G.}~\bibnamefont
  {Zumofen}}\ and\ \bibinfo {author} {\bibfnamefont {J.}~\bibnamefont
  {Klafter}},\ }\href {\doibase 10.1103/PhysRevE.47.851} {\bibfield  {journal}
  {\bibinfo  {journal} {Phys. Rev. E}\ }\textbf {\bibinfo {volume} {47}},\
  \bibinfo {pages} {851} (\bibinfo {year} {1993})}\BibitemShut {NoStop}%
\bibitem [{\citenamefont {Artuso}\ and\ \citenamefont
  {Cristadoro}(2003)}]{Artuso2003}%
  \BibitemOpen
  \bibfield  {author} {\bibinfo {author} {\bibfnamefont {R.}~\bibnamefont
  {Artuso}}\ and\ \bibinfo {author} {\bibfnamefont {G.}~\bibnamefont
  {Cristadoro}},\ }\href {\doibase 10.1103/PhysRevLett.90.244101} {\bibfield
  {journal} {\bibinfo  {journal} {Phys. Rev. Lett.}\ }\textbf {\bibinfo
  {volume} {90}},\ \bibinfo {pages} {244101} (\bibinfo {year}
  {2003})}\BibitemShut {NoStop}%
\bibitem [{\citenamefont {Thaler}(1983)}]{Thaler1983}%
  \BibitemOpen
  \bibfield  {author} {\bibinfo {author} {\bibfnamefont {M.}~\bibnamefont
  {Thaler}},\ }\href@noop {} {\bibfield  {journal} {\bibinfo  {journal} {Isr.
  J. Math.}\ }\textbf {\bibinfo {volume} {46}},\ \bibinfo {pages} {67}
  (\bibinfo {year} {1983})}\BibitemShut {NoStop}%
\bibitem [{\citenamefont {Cox}(1962)}]{Cox}%
  \BibitemOpen
  \bibfield  {author} {\bibinfo {author} {\bibfnamefont {D.~R.}\ \bibnamefont
  {Cox}},\ }\href@noop {} {\emph {\bibinfo {title} {Renewal theory}}}\
  (\bibinfo  {publisher} {Methuen},\ \bibinfo {address} {London},\ \bibinfo
  {year} {1962})\BibitemShut {NoStop}%
\bibitem [{\citenamefont {Feller}(1971)}]{Feller1971}%
  \BibitemOpen
  \bibfield  {author} {\bibinfo {author} {\bibfnamefont {W.}~\bibnamefont
  {Feller}},\ }\href@noop {} {\emph {\bibinfo {title} {An Introduction to
  Probability Theory and its Applications}}},\ \bibinfo {edition} {2nd}\ ed.,\
  Vol.~\bibinfo {volume} {2}\ (\bibinfo  {publisher} {Wiley, New York},\
  \bibinfo {year} {1971})\BibitemShut {NoStop}%
\bibitem [{\citenamefont {Barkai}\ and\ \citenamefont
  {Cheng}(2003)}]{Barkai2003b}%
  \BibitemOpen
  \bibfield  {author} {\bibinfo {author} {\bibfnamefont {E.}~\bibnamefont
  {Barkai}}\ and\ \bibinfo {author} {\bibfnamefont {Y.}~\bibnamefont {Cheng}},\
  }\href@noop {} {\bibfield  {journal} {\bibinfo  {journal} {J. Chem. Phys.}\
  }\textbf {\bibinfo {volume} {118}},\ \bibinfo {pages} {6167} (\bibinfo {year}
  {2003})}\BibitemShut {NoStop}%
\bibitem [{\citenamefont {Godr{\`e}che}\ and\ \citenamefont
  {Luck}(2001)}]{God2001}%
  \BibitemOpen
  \bibfield  {author} {\bibinfo {author} {\bibfnamefont {C.}~\bibnamefont
  {Godr{\`e}che}}\ and\ \bibinfo {author} {\bibfnamefont {J.~M.}\ \bibnamefont
  {Luck}},\ }\href@noop {} {\bibfield  {journal} {\bibinfo  {journal} {J. Stat.
  Phys.}\ }\textbf {\bibinfo {volume} {104}},\ \bibinfo {pages} {489} (\bibinfo
  {year} {2001})}\BibitemShut {NoStop}%
\bibitem [{\citenamefont {Dynkin}(1961)}]{Dynkin1961}%
  \BibitemOpen
  \bibfield  {author} {\bibinfo {author} {\bibfnamefont {E.}~\bibnamefont
  {Dynkin}},\ }\href@noop {} {\bibfield  {journal} {\bibinfo  {journal}
  {Selected Translations in Mathematical Statistics and Probability (American
  Mathematical Society, Providence)}\ }\textbf {\bibinfo {volume} {1}},\
  \bibinfo {pages} {171} (\bibinfo {year} {1961})}\BibitemShut {NoStop}%
\bibitem [{\citenamefont {Hopf}(1937)}]{Hopf1937}%
  \BibitemOpen
  \bibfield  {author} {\bibinfo {author} {\bibfnamefont {E.}~\bibnamefont
  {Hopf}},\ }\href@noop {} {\emph {\bibinfo {title} {Ergodentheorie}}}\
  (\bibinfo  {publisher} {Springer},\ \bibinfo {year} {1937})\BibitemShut
  {NoStop}%
\bibitem [{\citenamefont {Thaler}(2000)}]{Thaler2000}%
  \BibitemOpen
  \bibfield  {author} {\bibinfo {author} {\bibfnamefont {M.}~\bibnamefont
  {Thaler}},\ }\href@noop {} {\bibfield  {journal} {\bibinfo  {journal} {Studia
  Math}\ }\textbf {\bibinfo {volume} {143}},\ \bibinfo {pages} {103} (\bibinfo
  {year} {2000})}\BibitemShut {NoStop}%
\bibitem [{\citenamefont {Thaler}(2005)}]{Thaler2005}%
  \BibitemOpen
  \bibfield  {author} {\bibinfo {author} {\bibfnamefont {M.}~\bibnamefont
  {Thaler}},\ }\href@noop {} {\bibfield  {journal} {\bibinfo  {journal} {Stoch.
  Dyn}\ }\textbf {\bibinfo {volume} {5}},\ \bibinfo {pages} {425} (\bibinfo
  {year} {2005})}\BibitemShut {NoStop}%
\bibitem [{\citenamefont {Ignaccolo}\ \emph {et~al.}(2001)\citenamefont
  {Ignaccolo}, \citenamefont {Grigolini},\ and\ \citenamefont
  {Rosa}}]{Ignaccolo2001}%
  \BibitemOpen
  \bibfield  {author} {\bibinfo {author} {\bibfnamefont {M.}~\bibnamefont
  {Ignaccolo}}, \bibinfo {author} {\bibfnamefont {P.}~\bibnamefont
  {Grigolini}}, \ and\ \bibinfo {author} {\bibfnamefont {A.}~\bibnamefont
  {Rosa}},\ }\href {\doibase 10.1103/PhysRevE.64.026210} {\bibfield  {journal}
  {\bibinfo  {journal} {Phys. Rev. E}\ }\textbf {\bibinfo {volume} {64}},\
  \bibinfo {pages} {026210} (\bibinfo {year} {2001})}\BibitemShut {NoStop}%
\bibitem [{\citenamefont {Korabel}\ and\ \citenamefont
  {Barkai}(2009)}]{Korabel2009}%
  \BibitemOpen
  \bibfield  {author} {\bibinfo {author} {\bibfnamefont {N.}~\bibnamefont
  {Korabel}}\ and\ \bibinfo {author} {\bibfnamefont {E.}~\bibnamefont
  {Barkai}},\ }\href@noop {} {\bibfield  {journal} {\bibinfo  {journal} {Phys.
  Rev. Lett.}\ }\textbf {\bibinfo {volume} {102}},\ \bibinfo {pages} {050601}
  (\bibinfo {year} {2009})}\BibitemShut {NoStop}%
\end{thebibliography}%
\end{document}